\documentclass[journal]{IEEEtran}

\usepackage{graphicx}
\usepackage{epsf}
\usepackage{amsmath, amsfonts}
\usepackage{latexsym}
\usepackage{subfigure}
\usepackage{multirow}
\usepackage{multicol}
\usepackage[table]{xcolor}
\usepackage[hidelinks]{hyperref}

\begin{document}

\title{\fontsize{16}{10}\selectfont Mitigation of Human EMF Exposure in 5G Downlink}

\author{
Imtiaz Nasim and Seungmo Kim

\thanks{I. Nasim and S. Kim are with Department of Electrical and Computer Engineering, Georgia Southern University in Statesboro, GA, USA (e-mail: \{in00206, seungmokim\}@georgiasouthern.edu).}
}

\maketitle

\begin{abstract}
While research on communications at frequencies above 6 gigahertz (GHz) has been primarily confined to performance improvement, their potentially harmful impacts on human health are not studied as significantly. Most of the existing studies that paid attention to the health impacts above 6 GHz focused only on the uplink due to closer contact with a transmitter to a human body. In this letter, we present the human electromagnetic field (EMF) exposure in the downlink of Fifth-Generation Wireless Systems (5G). Moreover, we propose a downlink protocol that guarantees the EMF exposure under a threshold while keeping the data rate above the 5G requirements.
\end{abstract}

\vspace{0.2 in}

\begin{IEEEkeywords}
5G; above 6 GHz; Downlink; Human EMF exposure; SAR; PD
\end{IEEEkeywords}

\IEEEpeerreviewmaketitle

\section{Introduction}\label{sec_intro}
Recently, there are strong warnings from scientists around the world on the harmful impacts of exposure to EMFs on human health in wireless communications systems adopting more highly concentrated EMF energy, such as 5G \cite{5gappeal_sep17}.

As a solution for the skyrocketing bandwidth demand, 5G is expected to achieve higher data rates \cite{ericsson15_mobility}, which again will lead to the need for higher signal power at a receiver. A recent link budget study \cite{tap17} indicates that comparing data rates from 1 to 6 Gbps, the required received power grows from -65 to -37.5 dBm.

Moreover, 5G targets to operate at higher frequencies (e.g., 28 GHz \cite{amitava18}) due to advantages such as (i) availability of far wider bandwidths than today’s cellular networks, and (ii) possible design of larger numbers of miniaturized antennas to be placed in small dimensions, attributed to very small wavelengths \cite{jsac14}. Because such high frequencies enable high-gain directional antenna arrays, radiation energy is focused in certain directions, which can yield increased power deposition in the main lobe points towards the human body \cite{shrivastava17}.

In this context, this letter proposes a 5G downlink protocol that maximizes the data rate while guaranteeing the EMF exposure under a safe level. Contributions of this letter can be highlighted in comparison to state-of-the art as follows:

First, while the prior work studied the uplink only \cite{rappaport15}-\cite{colombi18}, this letter examines the human EMF exposure in the \textit{downlink} of 5G. Due to the higher signal powers and highly directional antennas, 5G can elevate levels of EMF emission in downlink as well as uplink.

Second, this letter \textit{proposes a downlink transmission protocol that mitigates the human EMF exposure}. In the protocol, for a user equipment (UE), the serving base station (BS) is selected among the ones whose maximum EMF emission level is below a threshold. (Our results show that a UE can experience EMF levels exceeding the threshold within certain distances from the BS.)

Third, this letter highlights that specific absorption rate (SAR) is a more effective metric than power density (PD) in evaluating impacts of wireless communications on human health. Agencies such as the Federal Communications Commission (FCC) \cite{fcc01} and the International Commission on Non-Ionizing Radiation Protection (ICNIRP) \cite{icnirp98} do not have guidelines on EMF exposure in terms of SAR at frequencies above 6 and 10 GHz, respectively, due to shallow penetration of EMF into a human body at such high frequencies. However, SAR can display more information as being able to express the level of EMF energy that is actually `absorbed' into a human body.

\begin{table}[t]
\scriptsize
\caption{Parameters for BS in 5G, 4G, and 3.9G}
\centering
\begin{tabular}{|c|c|c|c|c}
\hline 
\textbf{Parameter} & \multicolumn{3}{|c|}{\textbf{Value}}\\ \hline \hline
&\multicolumn{1}{|c|}{\cellcolor{gray!10}5G} & \multicolumn{1}{|c|}{\cellcolor{gray!10}4G} & \multicolumn{1}{|c|}{\cellcolor{gray!10}3.9G }\\
\hline 
Standard (3GPP) & Release 15 & Release 12 & Release 9\\
Carrier frequency & {28 GHz} & 2 GHz & 1.9 GHz\\
Path loss model & {UMi \cite{tr38901}} & {UMi \cite{tr36873}} & {UMi \cite{ts25996}}\\
Inter-site distance (ISD) & {200 m} & 200 m & 1 km\\
Sectors/site & {3} & 3 & 3 or 6\\
Bandwidth & {850 MHz} & 20 MHz & 20 MHz\\
Max antenna gain & {8 dBi/element} & 8 dBi/element &17 dBi\\
Transmit power & 21 dBm/element & 44 dBm  & 43 dBm \\
Antenna layout ($\lambda/2$) & 8$\times$8 & 4 & 4\\
Antenna height & 10 m & 10 m & 32 m\\ \hline
Duplexing & \multicolumn{3}{|c|}{Time-division duplexing (TDD)} \\
Transmission scheme & \multicolumn{3}{|c|}{Singler-user (SU)-MIMO} \\
UE noise figure & \multicolumn{3}{|c|}{7 dB}\\
Temperature & \multicolumn{3}{|c|}{290 K}\\ \hline
\end{tabular}
\label{table_parameters}
\end{table}

\section{System Model}\label{sec_model}
This letter adopts the downlink of the 3rd Generation Partnership Project (3GPP) Release 15 \cite{tr38901} (representing 5G). The analysis is compared to the downlinks of Releases 12 \cite{tr36873} and 9 \cite{ts25996} (representing 4G Long-Term Evolution (LTE) and 3.9G, respectively). The parameters for the three systems are summarized in Table \ref{table_parameters}.

Commonly for 5G, 4G, and 3.9G, this letter assumes a fully loaded network in order to understand the worst possible EMF exposure. Specifically, none of the three systems are supposed to adopt the power control nor adaptive beamforming, which can reduce the number of UEs that are being served at a certain time instant. The reason for such a worst-case assumption is to provide a `conservative' suggestion on human safety, which leaves some safety margin as discussed in \cite{colombi18}.

\subsection{5G and 4G}
In 5G and 4G, a network consists of 19 sites, each of which is composed of 3 sectors \cite{tr38901}\cite{tr36873}.

Among three path loss models--Rural Macro (RMa), Urban Macro (UMa), and Urban Micro (UMi), our analysis found that the other two models yielded lower SAR levels while UMi showed considerably higher SAR levels throughout a cell. For this reason, as indicated in Table \ref{table_parameters}, this letter limits the scope of investigation to UMi.

For a BS in 5G and 4G, the antenna element pattern combined in the elevation and azimuth planes is given by

\begin{align}
A\left(\phi,\theta\right) = \min \left(A_{a}\left(\phi\right)+A_{e}\left(\theta\right), A_m\right)\rm{~[dB]}
\label{eq_antenna}
\end{align}
where $\phi$ and $\theta$ are angles of a beam on the azimuth and elevation plane, respectively; the angle at which a 3-dB loss occurs is 65$^{\circ}$;  $A_m =30$ dB is a maximum attenuation (or a front-to-back ratio) \cite{tr38901}. Finally, an antenna gain is given by
\begin{align}
G\left(\phi,\theta\right)=G_{max} - A\left(\phi,\theta\right) \rm{~[dB]}
\label{eq_geometry_G_final}
\end{align}
where $G_{max}$ is a maximum antenna gain.

\subsection{3.9G}
For UMi, the radius of a cell is 500 m \cite{ts25996}. The beam radiation pattern for a BS in 3.9G is also constructed using (\ref{eq_antenna}) and (\ref{eq_geometry_G_final}). Values for parameters are different in 3.9G \cite{ts25996}: BS 3-dB angle is 35$^\circ$; $A_m$ is 23 dB; and the antenna gain is assumed omnidirectional.

\section{Analysis of Human EMF Exposure}\label{sec_analysis}
Biological effects of EMF depend on the level of energy absorbed into the human tissues. The depth of penetration into the human tisisues depends on the frequency and conductivity of the tissues \cite{shrivastava17}. As mentioned in Section \ref{sec_intro}, above 6 GHz where 5G will likely operate, safety guidelines \cite{fcc01}\cite{icnirp98} are defined in terms of PD due to the shallow penetration at such high frequencies.

However, recent studies found that PD is not as useful as SAR or temperature in assessment of EMF exposure since SAR can display the level of EMF energy that is actually `absorbed' in the body \cite{shrivastava17}\cite{rappaport15} while PD cannot. Furthermore, SAR is a more adequate metric than temperature since it can be directly calculated from PD, which is easier to calculate. Also, the effect of temperature is likely to be dispersed over the long distance in downlinks. Therefore, this letter selects SAR as the primary metric that measures the human EMF exposure level in 5G downlinks.

PD is defined as the amount of power radiated per unit volume at a distance \textit{d} \cite{wu15}, which is given by
\begin{align}\label{eq_pd_d}
\text{PD}\left(d\right) = \frac{\left|E\left(d\right)\right|^2}{\rho_0} \rm{~~[W/m^2]}
\end{align}
where $E\left(d\right)$ is the incident electric field's complex amplitude and $\rho_0$ is the characteristic impedance of free space. It can be rewritten by using the transmitter's parameters as
\begin{align}\label{eq_pd_phi}
\text{PD}\left(d, \phi\right) = \frac{P_T G_T\left(d, \phi\right)}{4 \pi d^2}
\end{align}
where $P_T$ is a transmit power; $G_T$ is a transmit antenna gain; $d$ is a BS-UE distance (m).

At high frequencies such as 28 GHz, most of the energy of a signal incident on human tissue is deposited into the thin surface of skin \cite{love16}. This can be expressed in terms of SAR, as a function of PD. SAR is defined as a measure of incident energy absorbed per unit of mass and time and thus quantifies the rate at which the human body absorbs energy from an electromagnetic field, which measures the power dissipated per body mass. The local SAR value at a point p measured in W/kg \cite{love16} can be expressed as
\begin{align}\label{eq_sar_p}
\text{SAR}\left(\text{p}\right) = \frac{\sigma\left|E\left(\text{p}\right)\right|^2}{\rho} \rm{~~[W/kg]}
\end{align}
where $\sigma$ is the conductivity of the material and $\rho$ is the density of the material. The SAR at a point on the air-skin boundary \cite{chahat12} can be written as a function of PD$(d,\phi)$ as
\begin{align}\label{eq_sar_phi}
\text{SAR}\left(d,\phi\right) = \frac{2\text{PD}\left(d, \phi\right) \left(1 - R^2\right)}{\delta \rho}
\end{align}
where $R$ is the reflection coefficient \cite{wu15}, $\rho$ is the tissue mass density (1 $\text{g}/\text{cm}^3$ is used), and $\delta$ is the skin penetration depth (10$^{\text{-3}}$ m is used) \cite{rappaport15}.

Note that $d$ and $\phi$ depend on the position of a UE in a cell. Therefore, in order to evaluate over all the possible UE positions in a cell, the SAR is calculated as an average over the area of a `sector' in a cell, which is given by
\begin{align}\label{eq_mean}
\mathbb{E}[\text{SAR}\left(\mathtt{x}_{ue}\right)] = \frac{1}{\left|\mathcal{R}_k^2\right|} \int_{\mathtt{x}_{ue}^{\left(k\right)} \in \mathcal{R}_k^2} \text{SAR}\left(\mathtt{x}_{ue}\right) d\mathtt{x}_{ue}
\end{align}
where $\mathcal{R}_k^2$ denotes a two-dimensional space representing a sector and thus $\left|\mathcal{R}_k^2\right|$ is the area of a sector; $\mathtt{x}_{ue}$ is position of a UE in an $\mathcal{R}_k^2$, which determines $d$ and $\phi$. Uniform distribution of UEs on each of the X- and Y-axis of each sector, $\mathcal{R}_k^2$, was considered.

\section{Proposed Protocol}\label{sec_proposed_protocol}
Now this letter proposes a novel 5G downlink protocol that guarantees the EMF exposure under a threshold. In the protocol, PD precedes received signal strength (RSS); for a UE, the serving BS with the highest RSS is selected but the selection occurs strictly among the ones with PDs under a threshold. The threshold can be flexibly chosen according to the environment in which the system operates. This letter chooses the threshold to be the safety guideline provided commonly by the FCC \cite{fcc01} and ICNIRP \cite{icnirp98}.

Specifically, in the BS selection for a UE, the PD in (\ref{eq_sar_phi}) is substituted with PD$'$, which is defined as the maximum PD among the ones below the threshold and can be formally written as
\begin{align}
\text{PD}' = \max_{i \in \mathcal{S}} \text{PD}_i,
\end{align}
where $\mathcal{S} = \{i {\rm{~}} | {\rm{~}} \text{PD}_i < \gamma\}$ represents a set of all the BSs that can serve downlink to the given UE, with $i$ and $\gamma$ denoting the index of each BS (i.e., the $i$th BS) in $\mathcal{S}$ and the threshold on PD, respectively.

\begin{figure}[t]
\centering\includegraphics[width = 0.8\linewidth]{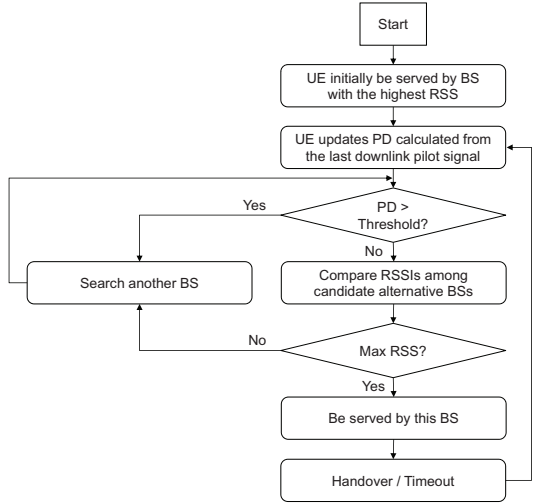}
\caption{Flowchart for the proposed protocol}
\label{fig_protocol_flowchart}
\vspace{-0.2 in}
\end{figure}

Fig. \ref{fig_protocol_flowchart} provides specifics for the operation flow of the proposed protocol. Each UE is initially served by the BS with the highest RSS, as in typical downlink protocols. However, the proposed protocol lets the UE update the PD as well, when it updates the information of the surrounding BSs for possible handovers. This update is accomplished via a downlink control channel--e.g., Physical Downlink Control Channel (PDCCH). Level of the PD received from a BS is used to examine whether it violates the threshold. If the current serving BS violates the threshold, the UE is handed over to the BS providing the next highest RSS, which is selected among the ones with PDs below the threshold. There are two scenarios that starts this entire procedure again. First, when UE is handed over to another BS, the new BS is selected via this process among $\mathcal{S}$. The second is a timeout. The protocol forces to periodically perform a BS search in case the current BS violates or set $\mathcal{S}$ changes.

The rationale that the proposed protocol operates in terms of PD is two-fold. First, keeping in mind that SAR is the primary metric for performance evaluation, a PD can always easily be converted to a SAR according to (\ref{eq_sar_phi}). Second, PD can be derived at a BS. If done at a UE, a separate feedback channel would be needed to report to the serving BS. However, exploiting the fact that a SAR is a function of PD, a BS can measure its PD via estimation of the distance to UE. This estimation can be performed by inferring the distance from the power of a received control signal--viz., via channel quality indicator (CQI) in Physical Uplink Control Channel (PUCCH). One benefit of this idea is that such `piggybacking' can reduce the feedback overhead between a UE and its serving BS, which finally yields a more efficient cellular networking.

\section{Numerical Results}\label{sec_results}
This section evaluates the performance of the proposed protocol (described in Section \ref{sec_proposed_protocol}) based on the EMF exposure analysis framework (described in Section \ref{sec_analysis}).

\begin{figure}
\centering
\begin{minipage}[t]{.329\textwidth}
\centering
\includegraphics[width = 1\textwidth]{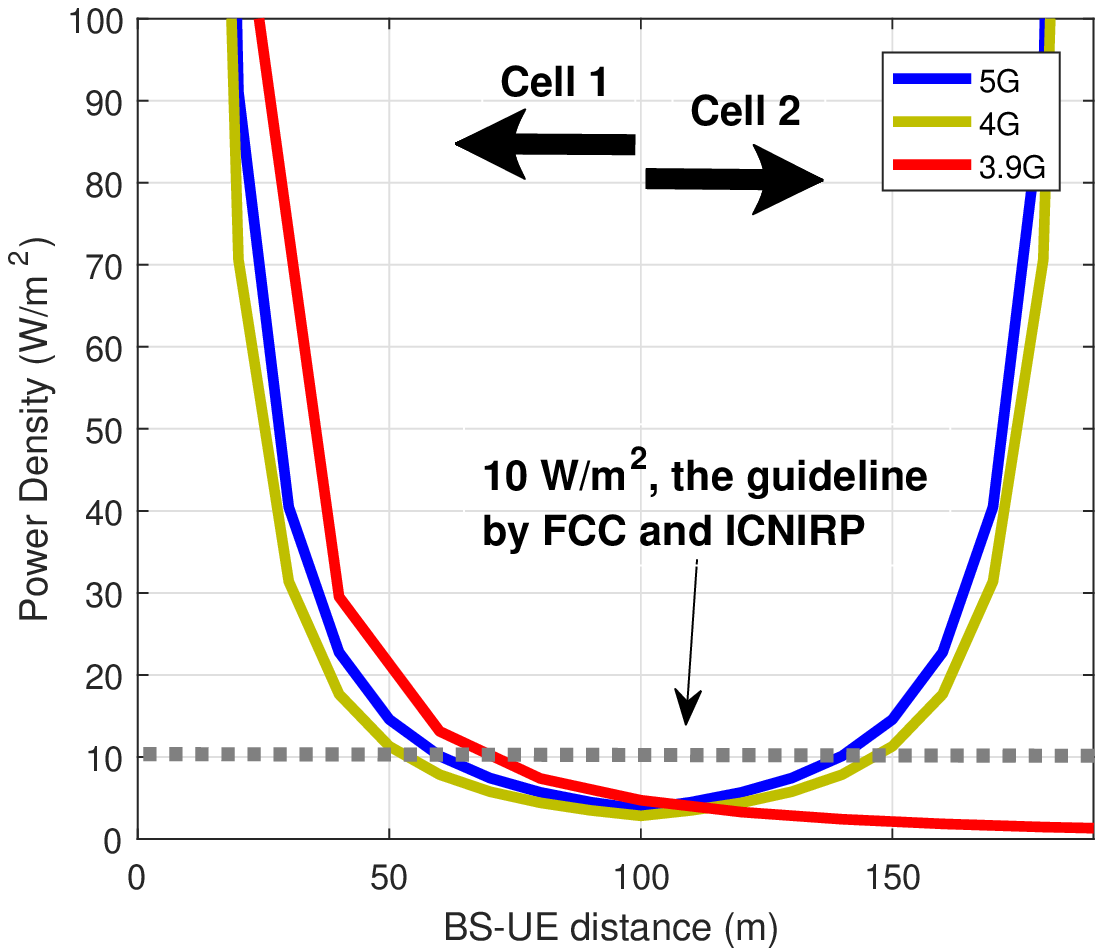}
\caption{PD versus BS-UE distance}
\label{fig_pd_mean}
\end{minipage}
\begin{minipage}[t]{.329\textwidth}
\centering
\includegraphics[width = 1\textwidth]{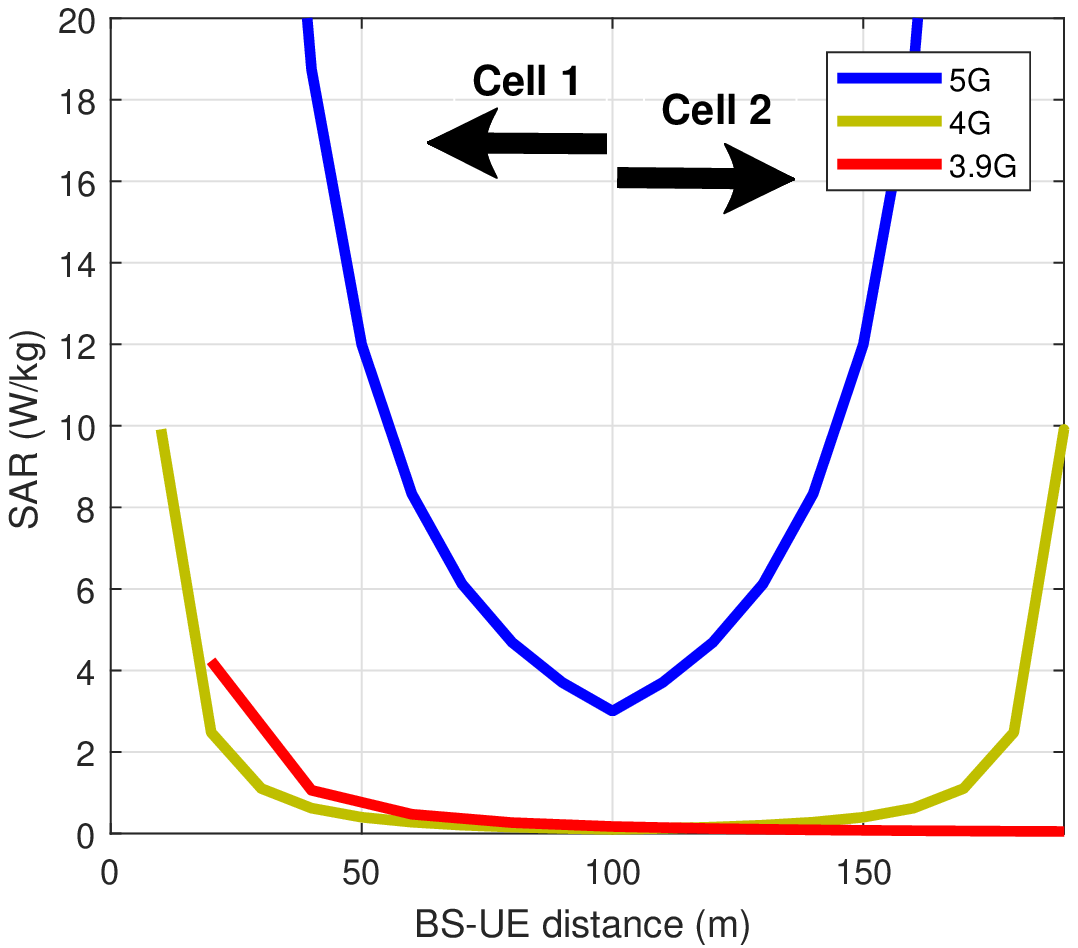}
\caption{SAR versus BS-UE distance}
\label{fig_sar_mean}
\end{minipage}
\vspace{-0.2 in}
\end{figure}

Figs. \ref{fig_pd_mean} and \ref {fig_sar_mean} demonstrate the levels of EMF exposure in terms of PD and SAR. In 5G and 4G, a BS is placed at 0 and 200 m, while 3.9G has its only BS at 0 m, according to the ISD of 100 and 500 m, respectively. (See Table \ref{table_parameters}.) Note that Fig. \ref{fig_pd_mean} calculates substitution of (\ref{eq_pd_phi}) into (\ref{eq_mean}), and Fig. \ref{fig_sar_mean} directly computes (\ref{eq_mean}). In order to consider geographic variation of the EMF exposure, these average quantities are calculated from 10$^4$ `drops,' each of which generates 10 UEs per sector. As mentioned in Section \ref{sec_model}, the system is assumed to be fully loaded; the calculation considers a time length that is enough for all the 10 UEs that are served based on time-division duplexing (TDD).

Fig. \ref{fig_pd_mean} shows that the PD for 5G falls below the guideline of 10 W/m$^2$, set by the FCC \cite{fcc01} and ICNIRP \cite{icnirp98}, with the BS-UE distance larger than 55 m. However, the problem starts to be seen when one looks at the same situation in terms of SAR.

Fig. \ref{fig_sar_mean} presents this problem. Note that PD is converted to SAR based on (\ref{eq_sar_phi}). The key observation is that 5G shows remarkably higher SARs throughout a cell compared to 4G and 3.9G, which suggests the necessity of an EMF mitigation scheme for the 5G. The problem is even more highlighted with consideration that the SAR guideline is set at 1.6 W/kg \cite{fcc01} with frequencies under 6 GHz.

The reason for this higher SAR is two-fold. First, in a 5G downlink, the \textit{shallower penetration} into human tissue yields higher level of absorption at the surface of human skin. This is confirmed from the fact that an instantaneous SAR is inverse-proportional to the penetration depth $\delta$ as shown in (\ref{eq_sar_phi}). Second, 5G presents several key differences in the system configuration. The \textit{small-cell topology} \cite{tr38901} is adopted in 5G to overcome the higher attenuation at high operating frequencies (i.e., 28 GHz and above), which yields the smaller BS-UE distance. Also, \textit{larger phased array antennas} contribute to the increase of PD in (\ref{eq_pd_phi}), which in turn results in higher SARs.

\begin{figure*}
\centering
\begin{minipage}[t]{.329\textwidth}
\centering
\includegraphics[width = 1\textwidth]{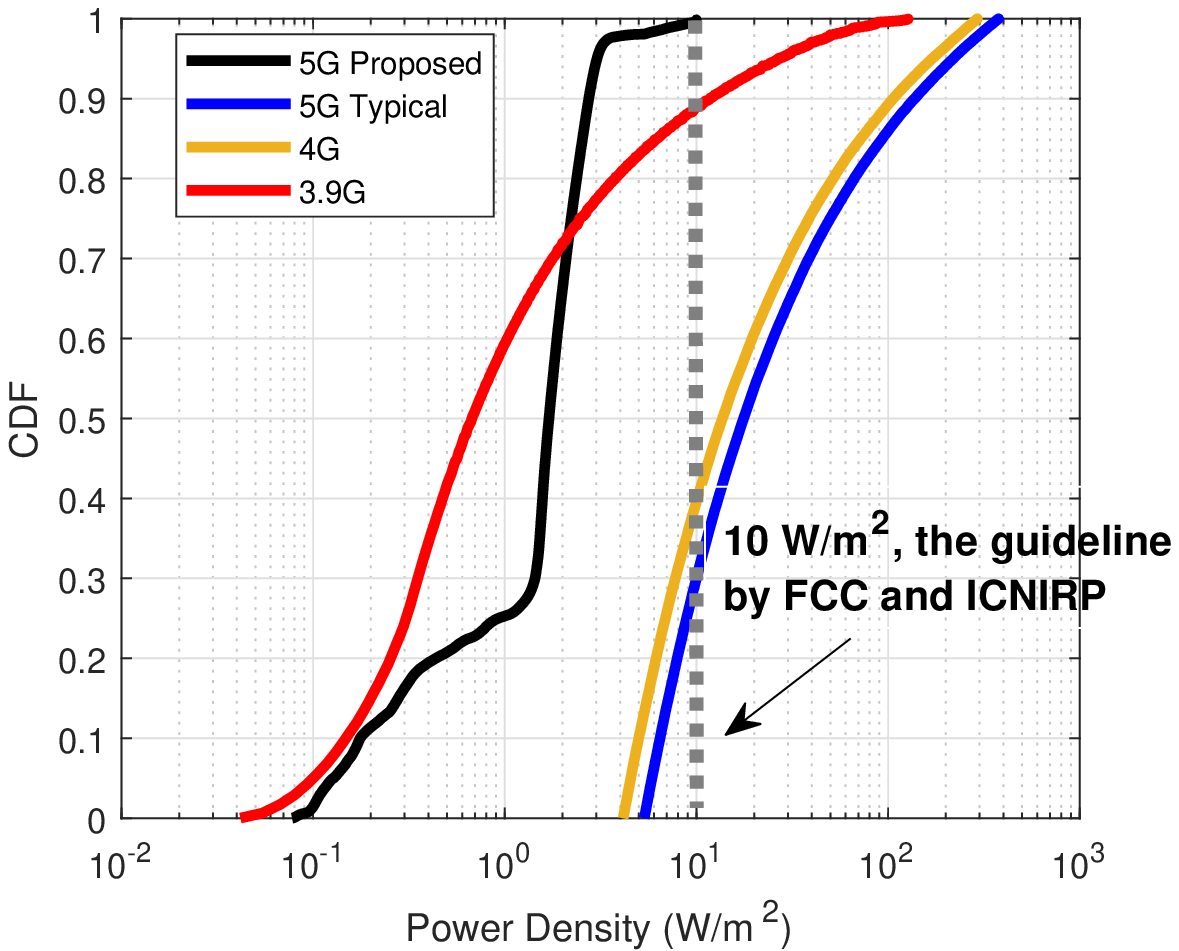}
\caption{Distribution of PD}
\label{fig_pd_cdf}
\end{minipage}
\begin{minipage}[t]{.329\textwidth}
\centering
\includegraphics[width= 1\textwidth]{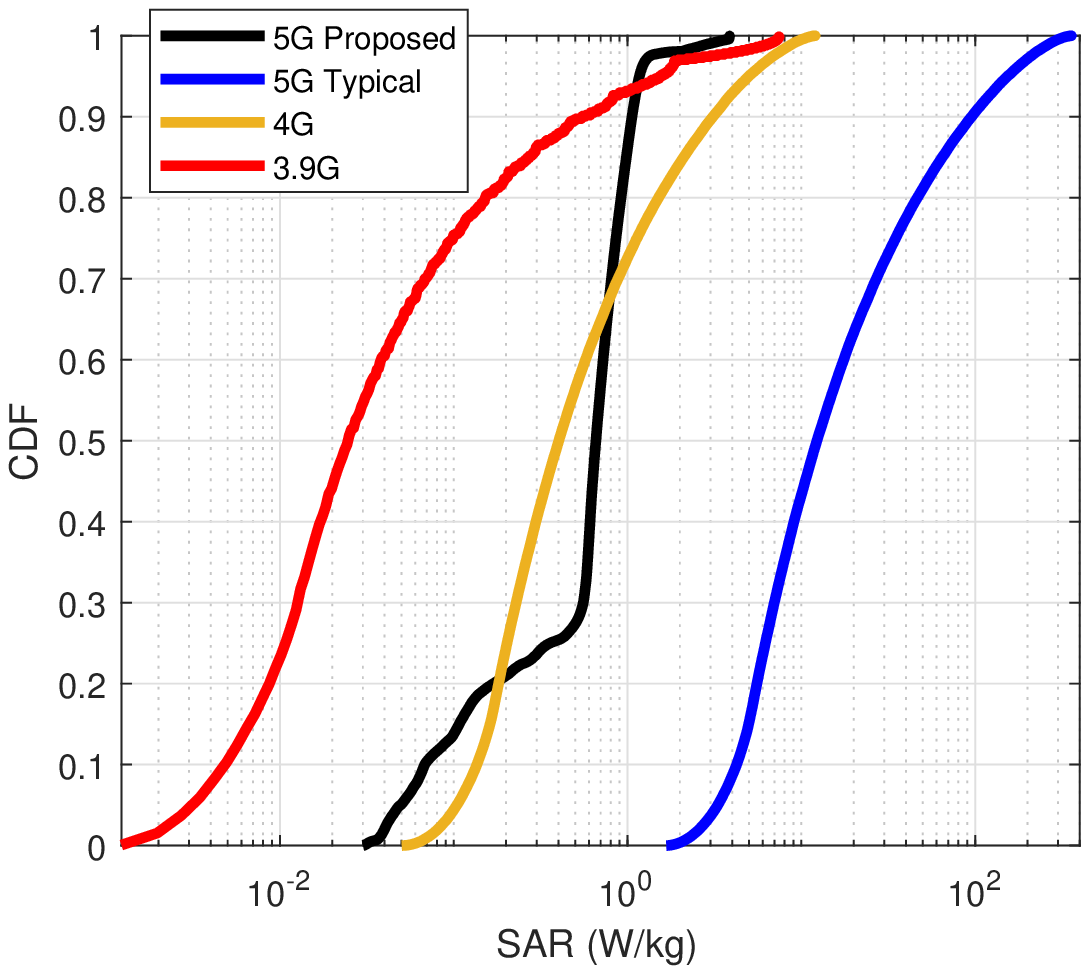}
\caption{Distribution of SAR}
\label{fig_sar_cdf}
\end{minipage}
\begin{minipage}[t]{.329\textwidth}
\centering
\includegraphics[width=1\textwidth]{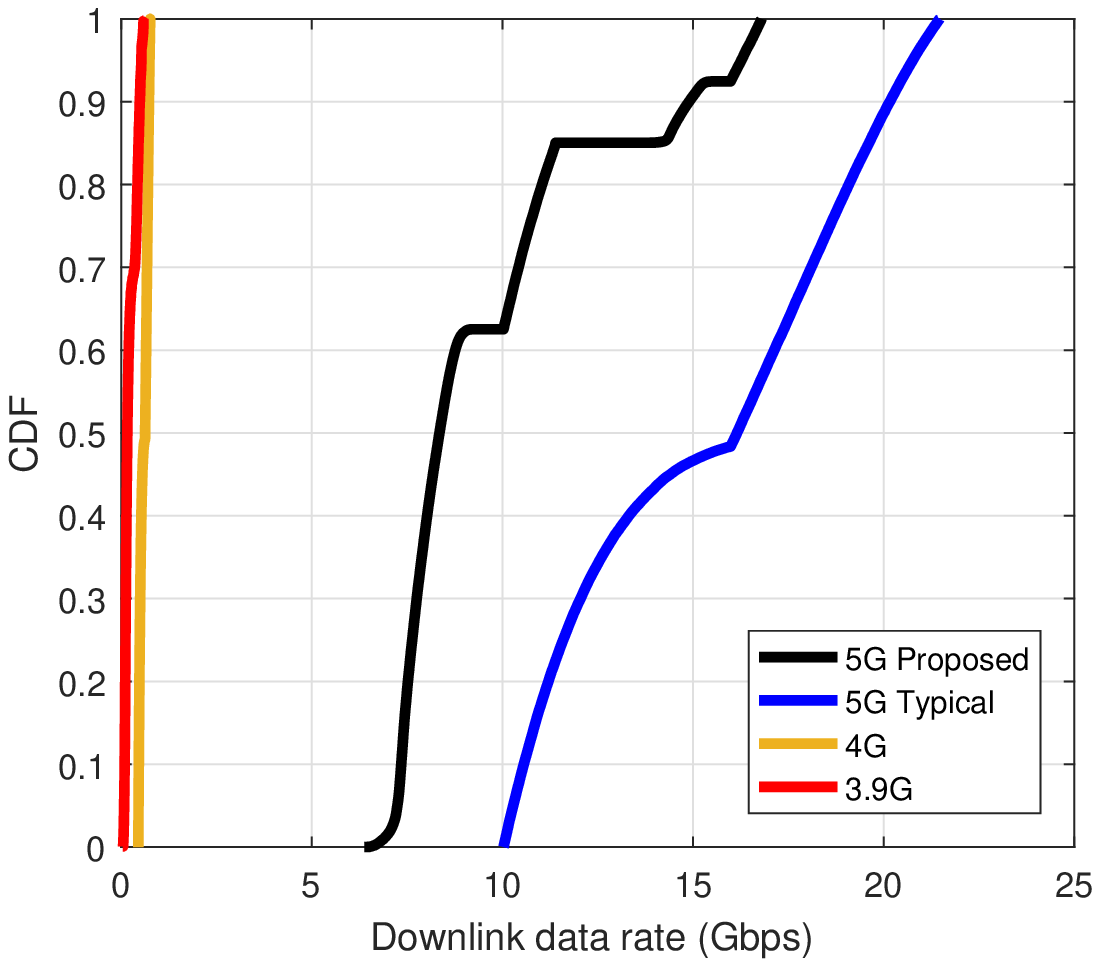}
\caption{Distribution of data rate}
\label{fig_rate_cdf}
\end{minipage}
\vspace{-0.2 in}
\end{figure*}

Now, the proposed protocol is applied to 5G and its impact is demonstrated. Figs. \ref{fig_pd_cdf} and \ref {fig_sar_cdf} show mitigation of the EMF exposure (in terms of PD and SAR, respectively) that the proposed protocol provides in the 5G downlink. The cumulative distribution function (CDF) is adopted in order to present a comprehensive view over an entire 19-cell layout.

Fig. \ref{fig_pd_cdf} presents that in 5G, nearly 70\% of the UEs in a cell are exposed to EMF levels higher than the guideline of 10 W/m$^2$. The proposed protocol sets the threshold at the guideline, which ensures that now PD never exceeds in the system. Although the threshold can be replaced with any updated one, the mitigation of PD that the proposed protocol introduces is well observed.

Derived from this tendency in PD, the very similar holds when represented in terms of SAR. Fig. \ref {fig_sar_cdf} suggests that our proposed protocol reduces the SAR level for 5G throughout the network. Although there is no guideline set in terms of SAR above 6 GHz, due to the fact that SAR is a more informative metric than PD, it is of significance to display the result in SAR. Further, we hope that this result can urge swift movement in setting up regulations for downlinks in terms of SAR at higher frequencies such as 28 GHz.

For a balanced study, this letter also provides the downside of the proposed protocol. Fig. \ref{fig_rate_cdf} shows the downlink data rate achieved throughout the 19-cell layout based on $B \log_2(1+\rm{SNR})$ where $B$ refers to the bandwidth, the value of which is provided in Table \ref{table_parameters}, and SNR indicates signal-to-noise ratio. In the proposed protocol, \textit{the downlink data rate is sacrificed} mainly due to selection of the serving BS only among the ones with PD under a threshold, which may exclude the BS with the highest RSS. Note that a latest practical study noted that expectation on the channel capacity in 5G is 7.73 bps/Hz \cite{amitava18}. As shown in Fig. \ref{fig_rate_cdf}, downlink data rates that are achieved by the proposed protocol range from 7 to 17 Gbps. Dividing this by the bandwidth of 850 MHz, the channel capacity is derived to range from 8.2353 to 20 bps/Hz. This suggests that the proposed protocol is still able to deliver sufficiently high data rates throughout a network.

\section{Conclusions}
Distinguished from the prior work that focused on human EMF exposure above in uplinks only, this letter investigated the significance of that in downlinks. Our results showed that the EMF exposure problem could be highlighted when displayed in terms of SAR, this letter proposes a downlink protocol that mitigates the human EMF exposure. The results show that the proposed protocol is effective in reduction of the human EMF exposure. It is also shown that the proposed protocol entails sacrifice in downlink data rates, but it still provides a sufficiently serviceable level of channel capacity.



\begin{thebibliography}{99}
\setlength{\parskip}{0.01em}

\bibitem{5gappeal_sep17} ``Scientists warn of potential serious health effects of 5G,'' Available at \url{https://ehtrust.org/wp-content/uploads/Scientist-5G-appeal-2017.pdf}

\bibitem{ericsson15_mobility} Ericsson, \textit{Ericsson mobility report on the pulse of the networked society}, Ericsson AB, Nov. 2015. Available at \url{http://www.ericsson.com/res/docs/2015/mobility-report/ericsson-mobility-report-nov-2015.pdf}

\bibitem{tap17} T. Tuovinen, N. Tervo, and A. Parssinen, ``Analyzing 5G RF system performance and relation to link budget for directive MIMO,'' \textit{IEEE Transactions on Antennas and Propagation}, vol. 65, iss. 12, Sep. 2017.

\bibitem{jsac14} M. Akdeniz, Y. Liu, M. Samimi, S. Sun, S. Rangan, T. Rappaport, and E. Erkip, ``Millimeter wave channel modeling and cellular capacity evaluation,'' \textit{IEEE J. Sel. Areas Commun.}, vol. 32, no. 6, 2014.

\bibitem{shrivastava17} P. Shrivastava and T. R. Rao, ``Specific absorption rate distributions of a tapered slot antenna at 60 GHz in personal wireless devices,''\textit{IEEE Antennas and Propagation Magazine}, vol. 59, Dec. 2017.

\bibitem{rappaport15} T. Wu, T. Rappaport, and C. Collins, ``The human body and millimeter-wave wireless communication systems: interactions and implications,'' in \textit{Proc. IEEE International Conference on Communications (ICC)}, 2015.


\bibitem{love16} M. Castellanos, D. Love, and B. Hochwald, ``Hybrid precoding for millimeter wave systems with a constraint on user electromagnetic radiation exposure,'' in \textit{Proc. Asilomar Conference on Signals, Systems and Computers}, Nov. 2016.

\bibitem{sambo15} Y. A. Sambo, F. Heliot, and M. Imran, ``Electromagnetic emission-aware scheduling for the uplink of coordinated OFDM wireless systems,'' in \textit{Proc. IEEE Online Conference on Green Communications (OnlineGreenComm)}, 2015.

\bibitem{colombi18} D. Colombi, B. Thors, C. Tornevik, and Q. Balzano, ``RF energy absorption by biological tissues in close proximity to millimeter-wave 5G wireless equipment,'' \textit{IEEE Access}, vol. 6, Aug. 2018.

\bibitem{fcc01} Federal Communications Commission, ``Evaluating Compliance with FCC guidelines for human exposure to radiofrequency electromagnetic fields,'' \textit{Supplement C Edition 01-01 to OET Bulletin 65 Edition 97-01}, Jun. 2001.

\bibitem{icnirp98} International Commission on Non-Ionizing Radiation Protection, ``ICNIRP guidelines: for limiting exposure to time-varying electric, magnetic and electromagnetic fields (up to 300 GHz),'' \textit{Health Physics}, vol. 74, no. 4, 1998.


\bibitem{tr38901} 3GPP TR 38.901, ``Channel model for frequencies from 0.5 to 100 GHz (Release 14),''  v14.3.0, Dec. 2017.

\bibitem{ts25996} 3GPP TR 25.996, ``Spatial channel model for multiple input multi output (MIMO) simulations (Release 9),''  v9.0.0, Dec. 2009.

\bibitem{tr36873} 3GPP TR 36.873, ``Study on 3D channel model for LTE (Release 12),'' v12.0.0, Sep. 2014.
Jun. 2015.

\bibitem{amitava18} A. Ghosh, ``5G New Radio (NR) : physical layer overview and performance,'' \textit{IEEE Communication Theory Workshop}, May 2018.

\bibitem{wu15} T. Wu, T. Rappaport, and C. Collins, ``Safe for generations to come: Considerations of safety for millimeter waves in wireless communications,'' \textit{IEEE Microwave Magazine}, vol. 16, no. 2, 2015.

\bibitem{chahat12} N. Chahat, M. Zhadobov, L. Le Coq, S. Alekseev, and R. Sauleau, ``Characterization of the interactions between a 60-GHz antenna and the human body in an off-body scenario,'' \textit{IEEE Trans. Antennas Prop.}, vol. 60, no. 12, 2012.


\end{thebibliography}
\end{document}